\documentstyle[12pt,fleqn]{article}
\def\soc{{\rm C}_{60}}
\def\rug{{\rm C}_{70}}
\def\nex{N_{\rm ex}}

\def\beeq{\begin{equation}}
\def\eneq{\end{equation}}
\def\beeqa{\begin{eqnarray}}
\def\eneqa{\end{eqnarray}}

\addtocounter{section}{-1}
\begin{document}

\begin{center}

\vspace{2in}

{\large {\bf{Doping Effects and Electronic States in C$_{\bf 60}$-Polymers
} } }

\vspace{1cm}

{\rm Kikuo Harigaya\footnote[1]{E-mail address: 
harigaya@etl.go.jp.}
}\\

\vspace{1cm}

{\sl Fundamental Physics Section,\\
Electrotechnical Laboratory,\\ 
Umezono 1-1-4, Tsukuba, Ibaraki 305, Japan}

\vspace{1cm}

(Received ~~~~~~~~~~~~~~~~~~~~~~~~~~~~~~~~~)
\end{center}

\Roman{table}

\vspace{1cm}

\noindent
{\bf Abstract}\\
Band structures of $\soc$-polymers are studied changing 
conjugation conditions and the electron number.  We use a 
Su-Schrieffer-Heeger type semiempirical model.  In the 
neutral $\soc$-polymer, electronic structures change among 
direct-gap insulator and the metal, depending on the degree 
of conjugations.  The $\soc$-polymer doped with one electron 
per one molecule is always a metal.  The energy difference 
between the highest-occupied state and the lowest-unoccupied 
state of the neutral system becomes smaller upon doping owing 
to the polaron effects.  The $\soc$-polymer doped with two 
electrons per one $\soc$ changes from an indirect-gap 
insulator to the direct-gap insulator, as the conjugations 
become stronger.



\pagebreak

\section{Introduction}

Recently, it has been found that the linear $\soc$-polymer
is realized in alkali-metal doped $\soc$ crystals: $A_1\soc$ 
($A=$K, Rb, Cs) [1-4], and much attention has been focused 
on their solid state properties.  One electron per one $\soc$ 
is doped in the polymer chain.  It seems that Fermi surfaces 
exist in high temperatures, but the system shows antiferromagnetic 
correlations in low temperatures than about 50K [1].  The 
structure of the $\soc$-polymer is displayed in Fig. 1(a).  The 
$\soc$ molecules are arrayed in a linear chain.  The bonds 
between $\soc$ are formed by the [2+2] cycloaddition mechanism.

Several calculations of the electronic structures have
been performed.  For example, a tight-binding calculation 
of a linear chain [5] has been reported, and the relation 
to the antiferromagnetic ground state has been discussed.  
A semiempirical tight-binding model [6] analogous to the 
Su-Schrieffer-Heeger (SSH) model [7] of conjugated polymers 
has been proposed, and the possibility of the charge density 
wave state has been pointed out.  The band calculation
by the first principle method has also been done [8].
The electronic structures can become three dimensional
when distances between $\soc$-polymer chains are short,
while they remain one dimensional when the distances are longer.

The above works have been focused upon electronic structures
of the $\soc$-polymer doped with one electron per one $\soc$.
The electronic structures may depend sensitively upon the 
conjugation conditions even in the neutral polymer, 
because the several bonds connecting neighboring molecules 
are largely distorted and the mixing between $\sigma$- and 
$\pi$-orbitals will change only by slight change of the bond 
structures [2].   We would like to study effects of the 
change of the conjugation conditions by introducing a 
phenomenological parameter in a tight-binding model.  The 
model is an extension of the SSH-type model which has been 
applied to $\soc$ [9,10] and $\rug$ [10,11] molecules.  
We look at electron densities, and band structures, in order 
to discuss metal-insulator changes and polaron effects.  We 
note that the calculations of the neutral system have been 
reported in [12].  We, however, include the results in 
order to discuss systematic variations from the neutral system 
to the doped systems.

The main conclusions are as follows: (1) In the neutral 
$\soc$-polymer, electronic structures change among direct-gap 
insulators and the metal, depending on the degree of 
conjugations.  The high pressure experiments may be able 
to change conjugation conditions in the chain direction, 
and the electronic structure changes could be observed.  
(2)  The $\soc$-polymer doped with one electron per one 
molecule is always a metal.  The energy difference between 
the highest-occupied state and the lowest-unoccupied state 
of the neutral system becomes smaller upon doping owing to 
the polaron effects.  (3) When the $\soc$-polymer is doped 
with two electrons per one $\soc$, the system is insulating.  
When the conjugation in the direction of the polymer 
chain is smaller, it is a direct-gap insulator.  The energy 
gap becomes indirect when the conjugation is stronger.

In the next section, our tight-binding model is introduced and
an idea of changing conjugation conditions is discussed.
The section 3 is devoted to numerical results.
The paper is closed with a short summary in section 4.

\section{Model}

We would like to apply an SSH-type model to a $\soc$-polymer.
In the previous works [9-11], we have proposed the extended 
SSH model to $\soc$ and $\rug$.  In $\soc$, all the carbon 
atoms are equivalent, so it is a good approximation to neglect 
the mixing between $\pi$- and $\sigma$-orbitals.  The presence 
of the dimerization and the energy level structures of the neutral
$\soc$ molecule can be quantitatively described by the calculations
within the adiabatic approximation.  In $\rug$, the molecular 
structure becomes longer, meaning that the degree of the mixing 
between $\pi$- and $\sigma$-characters are different depending 
on carbon sites.  In this respect, the extended SSH model does 
not take account of the difference of the mixings.  However, it 
has been found [10,11] that qualitative characters of the 
electronic level structures are reasonably calculated when the 
extended SSH model is applied to the $\rug$.  This is a valid 
approach because the energy positions of the $\sigma$-orbitals 
are deep enough and their effects can be simulated by the 
classical springs in the first approximation.

In this paper, we assume the same idea that the lattice structures 
and the related molecular orbitals of each $\soc$ molecule in the 
$\soc$-polymer can be described by the SSH-type model with one
orbital per one lattice site.  However, the mixings between the 
$\pi$- and $\sigma$-orbitals near the four bonds, $\langle i,j \rangle$ 
($i,j=1 - 4$), shown in Fig. 1(a) are largely different from those 
of regions far from the four bonds.  We shall shed light on this 
special character of bondings between neighboring $\soc$.  
Electronic structures would be largely affected by changes of 
conjugation conditions around the four bonds.  We shall introduce 
a semiempirical parameter $a$ as shown in the following 
Hamiltonian:
\beeqa
H_{\rm pol} &=&  a \sum_{l,\sigma} 
\sum_{\langle i,j \rangle = \langle 1,3 \rangle,
\langle 2,4 \rangle} (- t + \alpha y_{l,\langle i,j \rangle} )
( c_{l,i,\sigma}^\dagger c_{l+1,j,\sigma} + {\rm H.c.} )  \nonumber \\
&+&  (1-a) \sum_{l,\sigma} 
\sum_{\langle i,j \rangle = \langle 1,2 \rangle,
\langle 3,4 \rangle} (- t + \alpha y_{l,\langle i,j \rangle} )
( c_{l,i,\sigma}^\dagger c_{l,j,\sigma} + {\rm H.c.} ) \nonumber \\
&+& \sum_{l,\sigma} \sum_{\langle i,j \rangle = {\rm others}}
(- t + \alpha y_{l,\langle i,j \rangle} )
( c_{l,i,\sigma}^\dagger c_{l,j,\sigma} + {\rm H.c.} )  \nonumber \\
&+& \frac{K}{2} \sum_i \sum_{\langle i,j \rangle} y_{l,\langle i,j \rangle}^2,
\eneqa
where $t$ is the hopping integral of the system without the 
dimerization in the isolated $\soc$ molecule; $\alpha$ is the 
electron-phonon coupling constant which changes the hopping 
integral linearly with respect to the bond variable 
$y_{l,\langle i,j \rangle}$, where $l$ means the $l$th
molecule and $\langle i,j \rangle$ indicates the pair of
the neighboring $i$ and $j$th atoms; the atoms with $i=1 - 4$
are shown by numbers in Fig. 1(a) and the other $i$ within 
$5 \leq i \leq 60$ labels the remaining atoms in the same molecule;
$c_{l,i,\sigma}$ is an annihilation operator of the electron 
at the $i$th site of the $l$th molecule with spin $\sigma$; 
the sum is taken over the pairs of neighboring atoms; and the 
last term with the spring constant $K$ is the harmonic energy 
of the classical spring simulating the $\sigma$-bond effects.

As stated before, the parameter $a$ controls the strength of 
conjugations in the chain direction.  When $a=1$,
the bonding between atoms 1 and 2 (and also 3 and 4)
is completely broken and the orbitals would like $\pi$-orbitals
of the bonding between atoms 1 and 3 (and atomes 2 and 4).
The bond between the atoms 1 and 3 (and that between the atoms
2 and 4) becomes the classical double bond.   As $a$ becomes smaller, 
the conjugation between the neighboring molecule decreases
and the $\soc$ molecule becomes mutually independent.  In other
words, the interactions between molecules become smaller.
When $a=0$, the $\soc$ molecules are completely isolated
each other.  The band structures of the $\soc$ polymer will 
change largely depending on the conjugation conditions.  
This problem is the central issue of this paper.  We note that 
the operator at the lattice sites of the four membered rings is 
one of the relevant linear combinations of the effective $\sigma$-like 
components, assuming a possibility of local $\sigma$-conjugations 
at the four membered rings.  The similar assumption of the 
$\sigma$-conjugation has been used in Si-based polymers, 
for example, in [13].  We, however, use the term 
``conjugation" for simplicity in this paper, because the 
local $\sigma$-conjugations can be regarded as a part of the 
global conjugations which are extended over the system.

In the literature [6], the possibility of charge density 
wave states has been taken into account by regarding two molecule 
pair as a unit cell.  In contrast, our interests are focused 
on the one-dimensional band structure in the spatially 
homogeneous system, so we do not consider the doubled unit 
cell.  The present unit cell consists of one $\soc$ molecule.   
Using the lattice periodicity, we skip the index $l$ of the 
bond variable $y_{l,\langle i,j \rangle}$.  In other words, 
all the molecules in the polymer are assumed to have the same 
lattice structure.  The bond variables are determined by using 
the adiabatic approximation in the real space.  The same 
numerical iteration method as in [10] is used here.  We will 
change the parameter, $a$, within $0 \leq a \leq 1.0$, and the 
excess electron number per one $\soc$, $\nex$, is varied as 
$\nex = 0$, 1, and 2.  The other parameters, $t=2.1$eV, 
$\alpha = 6.0$eV/\AA, and $K = 52.5$eV/\AA$^2$, give the energy 
gap 1.904eV and the difference between the short and bond 
lengths 0.04557\AA~ for a $\soc$.  We shall use the same 
parameter set here.

\section{Numerical Results}

Figures 1(b), (c), and (d) show the excess electron distribution 
for the three conjugation conditions, $a=0.5$, 0.8, and 1.0.  
The labels of sites, A-I, are shown in Fig. 1(a).  Due to the 
reduced symmetry of the polymer chain, mutually symmetry equivalent 
sites have the same electron density.  The each label represents 
the site with the different electron density.  The white bars 
are for the case $\nex = 1$ and the black bars are for $\nex=2$.  
The excess electron density at the sites A is the largest for all 
the displayed cases.  The bond alternation patterns are largely 
distorted near these sites, so the electron density change is 
the largest too.  In Fig. 1(b), the densities at sites, D, F, 
and H, are relatively larger.  In Figs. 1(c) and (d), the densities 
are larger at sites D and I.  In this way, the sites, where 
excess electrons prone to accumulate and thus the dimerization 
patterns are highly distorted, are spatially localized in the 
molecular surface.  This is one of the polaron effects, which 
we have discussed in [10].  Here, we do not show dimerization 
patterns for simplicity.  We only note that the distortion of the 
dimerization is larger where the change of the electron density 
is larger.  The polaronic distortion pattern is different
from that in the isolated $\soc$, and this is owing to the
difference in the symmetry group.  Tanaka et al [5] have
drawn a schematic figure where the electron density change
is the largest at the molecule center, but the present result 
does not agree with this feature.  Numerical calculations
are necessary in order to derive the actual distributions.
They have not calculated the electron distributions in [5].

Next, we discuss band structures of electrons in detail.
Figures 2, 3, and 4 display the band structures for the conjugation
parameters, $a=0.5$, 0.8, and 1.0, respectively.  Figures (a),
(b), and (c) are for $\nex=0$, 1, and 2, respectively.  In each
figure, the unit cell is taken as unity, so the first Brilloune
zone extends from $-\pi$ to $\pi$.  Due to the inversion symmetry,
only the wavenumber region, $0 \leq k \leq \pi$, is shown in the figures.
We note that the band structures of the neutral system have been
reported in [12].  They are shown again in order to discuss
doping effects.

Figures 2(a), (b) and (c) show the band structures of the polymer for
$a=0.5$ and with $\nex=0$, 1, and 2, respectively.  In Fig. 2(a),
the highest fully occupied band is named as ``HOMO", and the lowest 
empty band as ``LUMO".  There is an energy gap about 0.8 eV at the 
zone center.  The system is a direct gap insulator.  When doped 
with one electron per $\soc$, the system is a metal as shown
by the presence of the Fermi surface in Fig. 2(b).  The system
is an insulator again when $\nex=2$, as shown in Fig. 2(c).
Here, the energy gap is at the boundary of the Brilloune zone,
i.e., at $k = \pi$.

As increasing the parameter $a$, the overlap of the HOMO band 
and LUMO band appears in the neutral system.  This is shown 
for $a=0.8$ in Fig. 3(a).  There are Fermi surfaces, so the 
system changes into a metal.  If $a$ increases further, the 
positions of the previous HOMO band and LUMO band are reversed 
as shown for $a = 1.0$ in Fig. 4(a).  The system becomes a 
direct gap insulator again.  The energy gap is at $k=\pi$.

When $\nex = 1$, the system is always a metal when $a$ varies.
The representative cases, $a = 0.8$ and 1.0, are displayed in
the Figs. 3(b) and 4(b).  The number of the Fermi surface is
two or four, depending upon the parameter $a$.  However, the 
metallic property is obtained for all the $a$ we take.
We also find that the HOMO band and LUMO band of the neutral
system shift into the energy gap upon doping.  The positions
of the other energy bands do not change so largely.
This is due to the polaronic distortion of the lattice,
which we have discussed in the calculation of an isolated
molecule [9,10].

When $\nex = 2$, the system turned out to be always an
insulator.  For smaller $a$, for example, $a = 0.5$ and 0.8,
the energy gap appears at $k=\pi$.  For larger $a$, for example,
$a = 1.0$ [Fig. 4(c)], the energy gap becomes an indirect gap.
The polaronic distortion becomes larger as the doping 
concentration increases.  Thus, the intrusions of the HOMO 
and LUMO bands of the neutral system become larger, too.

The above variations of the energy gap are summarized 
for the cases $\nex = 0$ and 2, where a finite energy 
gap appears for a certain $a$ value.  The results are 
shown in Figs. 5(a) and (b).  The white (black) squares 
indicate that the system is a direct gap insulator where 
there is a energy gap at $k=0$ ($\pi$).  The squares with 
the plus mean that the system is an indirect gap insulator.  
The crosses are for metals.  In the neutral system 
$\nex = 0$ [Fig. 5(a)], the energy gap decreases almost 
linearly for smaller $a$.  The system changes into a metal 
as $a$ increases, and finally an energy gap appears again.   
As has been discussed in [2], the conjugations 
between the bonds, $\langle 1,2 \rangle$ and $\langle 3,4 
\rangle$, might be weak.  So, we can assume that the larger 
parameter $a$ is reasonable for the real $\soc$-polymer.
There would be a good possibility that the realistic $a$ is 
in the region where we can expect metallic and insulating 
behaviors.  Therefore, it would be interesting to do 
experiments which give a high pressure to neutral systems 
in order to change conjugation conditions.  For 
$\nex = 2$ shown in Fig. 5(b), the system is a direct gap 
insulator with the energy gap at $k = \pi$ up to $a \sim 0.9$.  
The system turns into an indirect gap insulator near $a = 1.0$.

\section{Summary}

We have studied the variations of the band structures of 
the $\soc$-polymer.  We have changed conjugation 
conditions and the electron number.  A semiempirical model 
with SSH-type electron-phonon interactions has been proposed.  
Electron densities and band structures have been shown 
extensively, in order to discuss metal-insulator 
changes and polaron effects.  A possibility of observing
electronic structure changes in high pressure experiments
has been pointed out.

Recently, the other new phases of $\soc$ systems have been 
reported [14-16].  They are synthesized at high pressures.   
Two dimensional polymer structures [15] have been proposed 
and a tight-binding calculation [17] has been already 
reported.  The work which extends the present calculations 
to the novel two dimensional structures has been reported 
separately [18].

\pagebreak
\begin{flushleft}
{\bf References}
\end{flushleft}

\noindent
$[1]$ O. Chauvet, G. Oszl\`{a}nyi, L. Forr\'{o}, P. W. Stephens,
M. Tegze, G. Faigel, and A. J\`{a}nossy,
Phys. Rev. Lett.  72 (1994) 2721.\\
$[2]$ P. W. Stephens, G. Bortel, G. Faigel, M. Tegze,
A. J\`{a}nossy, S. Pekker, G. Oszlanyi, and L. Forr\'{o},
Nature 370 (1994) 636.\\
$[3]$ S. Pekker, L. Forr\'{o}, L. Mihaly, and A. J\`{a}nossy,
Solid State Commun.  90 (1994) 349.\\
$[4]$ S. Pekker, A. J\`{a}nossy, L. Mihaly, O. Chauvet,
M. Carrard, and L. Forr\'{o}, Science 265 (1994) 1077.\\
$[5]$ K. Tanaka, Y. Matsuura, Y. Oshima, T. Yamabe,
Y. Asai, and M. Tokumoto, Solid State Commun. 93 (1995) 163.\\
$[6]$ P. R. Surj\'{a}n and K. N\'{e}meth, Solid State Commun.
92 (1994) 407.\\
$[7]$ W. P. Su, J. R. Schrieffer, and A. J. Heeger,
Phys. Rev. B  22 (1980) 2099.\\
$[8]$ S. C. Erwin, G. V. Krishna, and E. J. Mele, Phys. Rev. B 
51 (1995) 7345.\\
$[9]$ K. Harigaya, J. Phys. Soc. Jpn. 60 (1991) 4001.\\
$[10]$ K. Harigaya, Phys. Rev. B 45 (1992) 13676.\\
$[11]$ K. Harigaya, Chem. Phys. Lett. 189 (1992) 79.\\
$[12]$ K. Harigaya, Phys. Rev. B 52 (1995) 7968.\\
$[13]$ T. Hasegawa, Y. Iwasa, H. Sunamura, T. Koda, Y. Tokura,
H. Tachibana, M. Matsumoto, and S. Abe, Phys. Rev. Lett.
69 (1992) 668.\\
$[14]$ Y. Iwasa, T. Arima, R. M. Fleming, T. Siegrist, O. Zhou,
R. C. Haddon, L. J. Rothberg, K. B. Lyons, H. L. Carter Jr., A. F. Hebard,
R. Tycko, G. Dabbagh, J. J. Krajewski, G. A. Thomas, and T. Yagi, Science 
264 (1994) 1570.\\
$[15]$ O. B\'{e}thoux, M. M\'{u}\~{n}ez-Regueiro, L. Marques, 
J. L. Hodeau, and M. Perroux, Paper presented at the Materials 
Research Society meeting, Boston, USA, November 29-December 3, 
1993, G2.9.\\
$[16]$ M. M\'{u}\~{n}ez-Regueiro, L. Marques, J. L. Hodeau,
O. B\'{e}thoux, and M. Perroux, Phys. Rev. Lett.
74 (1995) 278.\\
$[17]$ C. H. Xu and G. E. Scuseria, Phys. Rev. Lett. 74 (1995) 274.\\
$[18]$ K. Harigaya, Chem. Phys. Lett. 242 (1995) 585.\\

\pagebreak

\begin{flushleft}
{\bf Figure Captions}
\end{flushleft}

\mbox{}

\noindent
Fig. 1. (a) The crystal structure of the $\soc$ polymer.
The labels, A-I, indicate carbon atoms whose charge densities
are not equivalent due to the symmetry.  These labels are
used in (b), (c), and (d).  The conjugations along four bonds, 
which connect carbon atoms with labels, 1-4, are controled
by the parameter $a$ in eq. (1).  And, the excess electron 
distribution is shown for the three conjugation conditions, 
(b) $a=0.5$, (c) 0.8, and (d) 1.0.  The white bars are for 
the case $\nex = 1$ and the black bars are for $\nex=2$.

\mbox{}

\noindent
Fig. 2.  Band structures of the $\soc$-polymer of the case $a = 0.5$.
The excess electron number per one $\soc$ is (a) 0, (b) 1, 
and (c) 2, respectively.  In (a), the highest fully occupied band is 
named as ``HOMO", and the lowest empty band as ``LUMO".
The lattice constant of the unit cell is taken as unity.

\mbox{}

\noindent
Fig. 3.  Band structures of the $\soc$-polymer of the case $a = 0.8$.
The excess electron number per one $\soc$ is (a) 0, (b) 1, 
and (c) 2, respectively.  The lattice constant of the unit cell 
is taken as unity.

\mbox{}

\noindent
Fig. 4.  Band structures of the $\soc$-polymer of the case $a = 1.0$.
The excess electron number per one $\soc$ is (a) 0, (b) 1, 
and (c) 2, respectively.  In (a), the highest fully occupied band is 
named as ``HOMO", and the lowest empty band as ``LUMO".
The lattice constant of the unit cell is taken as unity.

\mbox{}

\noindent
Fig. 5.  The variations of the energy gap plotted against $a$.
The cases $\nex = 0$ and 2 are shown in (a) and (b), respectively.
The white (black) squares indicate that the system is
a direct gap insulator where there is a energy gap at
$k=0$ ($\pi$).  The squares with the plus symbol mean that the system
is an indirect gap insulator.  The crosses are for metallic cases.

\end{document}